\documentclass[aps,reprint,twocolumn,amsmath,amssymb,floatfix,showpacs]{revtex4}
\usepackage{graphicx}
\usepackage{dcolumn} 
\usepackage{bm}      
\usepackage[usenames,dvipsnames]{color}
\usepackage{ulem}    
\definecolor{orange}{cmyk}{0,0.5,1,0}

\def\ADD#1{#1}

\begin{document}

\title{On the emergence of helicity in rotating stratified 
       turbulence}
\author{Raffaele Marino$^1$, Pablo D. Mininni$^{1,2}$, Duane Rosenberg$^1$ and 
        Annick Pouquet$^1$}
\affiliation{
$^1$\ Institute for Mathematics Applied to Geosciences (IMAGe), 
                 CISL/NCAR, P.O. Box 3000, Boulder, Colorado 80307-3000, USA. \\
$^2$\ Departamento de F\'\i sica, Facultad de Ciencias Exactas y
                 Naturales, Universidad de Buenos Aires and IFIBA, CONICET, 
                 Ciudad Universitaria, 1428 Buenos Aires, Argentina.}
\date{\today}

\begin{abstract}
We perform numerical simulations of decaying rotating stratified turbulence and show, in the Boussinesq framework, that helicity (velocity-vorticity correlation), as observed in super-cell storms and hurricanes, is spontaneously created due to \ADD{an interplay between buoyancy and rotation} common to large-scale atmospheric and oceanic flows. Helicity emerges from  the joint action of eddies and of inertia-gravity waves (with inertia and gravity with respective associated frequencies $f$ and $N$), and it occurs when the waves are sufficiently strong. For $N/f < 3$ the amount of helicity produced is correctly predicted by a \ADD{quasi-}linear balance equation. Outside this regime, and up to the highest Reynolds number obtained in this study, namely $Re\approx 10000$, helicity production is found to be persistent for $N/f$ as large as $\approx 17$, and for $ReFr^2$ and  $ReRo^2 $ respectively as large as $\approx 100$ and $\approx 24000$.
\end{abstract}

\maketitle
\section{Introduction}

Symmetry breaking is a fundamental concept which has been quite fruitful in many physical applications \cite{anderson}. For a fluid the simplest way to break symmetry is to introduce helicity. In that case, the velocity covariance tensor is still expressible in terms of the magnitude of the distance between points, i.e., the fluid still has isotropic statistics, but mirror symmetry is broken whereby the covariance matrix has an anti-symmetric component which can be shown to be proportional to the total helicity $H_V=\left< {\bf u} \cdot {\bm \omega} \right> $, which is defined as the correlation between the velocity ${\bf u}$ and its curl, the vorticity ${\bm \omega}$. Helical structures abound in nature, from macroscopic organisms to elastomers; helical structures can cause erosion in river bends \cite{forterre}, and alter nutrient mixing properties in estuaries, in particular when interacting with tidal flows \cite{marine}. Helical flows are  observed as well in the atmosphere, in dust devils, tornadoes and  hurricanes \cite{moli}.

Helicity is an invariant of motion of the non-dissipative fluid equations involving the topology of field lines  \cite{Moffatt92}, including in the presence of rotation, but not with stratification, where instead potential vorticity is invariant and is essential in determining structures such as sharp jets in planetary atmospheres \cite{mcintyre}. Invariants are known to play a fundamental role in turbulence, since the nonlinear terms have to preserve such invariants at the level of triadic interactions in the incompressible case. However, helical (corkscrew) motions do not seem to alter the dynamics of homogeneous isotropic turbulence in the absence of both rotation and stratification, with the kinetic energy and helicity spectral densities (respectively $E_V(k)$ and $H_V(k)$, with $\int E_V(k)dk = \frac{1}{2}\left<u^2 \right>$ and $\int H_V(k)dk =H_V$) both following a Kolmogorov spectrum. This implies a slow $\sim 1/k$ decay of the relative helicity in Fourier space
$$\hat \Sigma(k) = \frac{ H_V(k)}{kE_V(k)} \ ,$$ 
with $\sigma_V({\bf x})=\cos ({\bf u}, {\bm \omega})$ the degree of alignment between velocity and vorticity in configuration space. However, it is straightforward to show that helicity is created point-wise by the alignment of vorticity and pressure  or shear gradients \cite{matthaeus}, and it is observed to be strong ($\sigma_V \sim \pm 1$) in the vortex filaments that are ubiquitous in isotropic fluid turbulence at small scale.

Invariants are also the stepping stone to determine inertial range behavior in turbulent flows; this principle is at the basis of statistical mechanics that has proven useful in predicting, for example, the inverse cascade of energy for a two-dimensional fluid \cite{kraichnan_montgomery}. In fact, a recent direct numerical simulation of the ideal three-dimensional fluid equations in the absence of waves, showed that, at intermediate times, a Kolmogorov spectrum develops at large scale, the effective dissipation for the large-scale fluid being produced by the eddy viscosity emanating from the small-scale equilibrated modes \cite{brachet}. However, non-conserved quantities can also play an important role through other mechanisms such as interactions with waves and large-scale hydrostatic and geostrophic balance \cite{Majda}.

When the fluid is conducting, magnetic helicity is an invariant in the ideal case and is central to minimum energy equilibria in plasmas such as in spheromaks, or in solar coronal mass ejections \cite{demoulin}. It is also known that the generation of large-scale magnetic fields occurs due to small-scale mechanic helicity $H_V$, and that in the presence of both rotation and stratification, helicity is created and thus a dynamo is facilitated in a wide variety of astrophysical settings \cite{branden_rev}. In the context of this work, it is important to note that although in this case the mechanic helicity is not an invariant any longer, it still plays an essential role in determining the scaling of the fields at large scales.

Rotating stratified turbulence  is important in the atmosphere and oceans, playing a crucial role in their dynamics. In the presence of waves, advective nonlinear interactions responsible for the complexity of turbulent flows have to compete with the waves and an equilibrium can be reached at some scale and broken at others, the best known example perhaps being the difference between the Garret-Munk and the Phillips spectra in internal waves in lakes or the oceans \cite{garrett, polzin}, where wave coupling in resonant triads leads  to mixing (like in coastal currents \cite{zeitlin}), to vertical dispersion \cite{lindborg}, and to enhanced dissipation \cite{fer}. A particular set of modes plays a major role, in the so-called slow manifold  for which the frequency of the waves tends to zero, and only turbulent interactions and standing potential vorticity modes remain. When rotation (only) is present in the fluid, strong relative helicity can alter the scaling of the distribution of energy among scales and lead to the occurrence of helical long-lived structures \cite{1536}.

What happens when stratification is also included? In the remainder of this paper, we address the question of rotating stratified turbulence  in the absence of 
forcing, as studied for example in \cite{bartello1, bartello2}, but concentrating on the creation of helicity and on the link between the evolution of helicity and \ADD{the balance of forces such as rotation, stratification, inertia, and pressure gradients, at scales large enough that the effect of nonlinearities (inertia) is small for strong waves. In the limit of zero nonlinearities, the resulting geostrophic balance} is crucial for weather forecasting and simulations of climate change. However, the consequences of \ADD{the interplay between rotation and stratification}, as far as helical motions are concerned, have been mostly ignored except for the pioneering work of Hide \cite{moffatt_bk}. In spite of this, helicity was hypothesized to be important in the atmosphere in the dynamics and persistence of rotating convective storms \cite{lilly} on the basis of the weakening of non-linear interactions in the so-called Lamb vector ${\bf u} \times {\bm \omega}$. 

It is also interesting that helicity is measured in the context of forecasting storms and tropical tornadoes, in particular in the presence of strong shear, and it can be used as an indicator of storm occurrence \cite{marko}. Note that it has been shown that in some cases (using a specific fully helical Beltrami forcing function), shear is created at large-scale in a rotating flow \cite{sen}.

Since helicity in rotating and stratified flows is no longer an invariant even in the absence of dissipation, its presence in these atmospheric storms can be accounted for but the physical mechanisms governing its creation, and the structures associated with it, remain unclear. In this paper, we perform a parametric study using direct numerical simulations in which we vary both rotation and stratification. In that framework, we show that a strongly rotating stratified flow can spontaneously create helicity at large scales.
 
\section{Equations and numerical procedure} 
\subsection{ Boussinesq equations and parameters}

\begin{figure*}
\includegraphics[width=8cm]{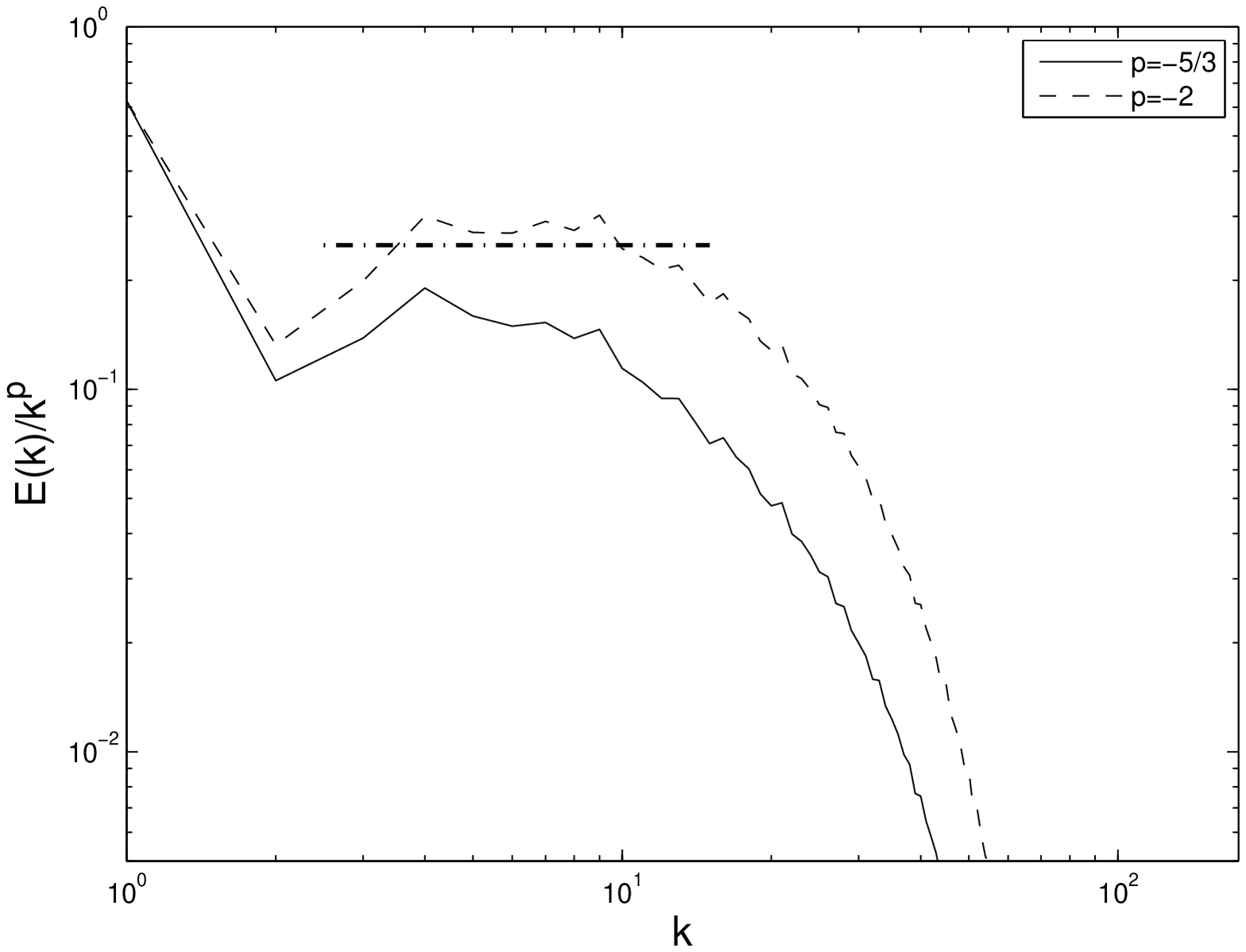} 
\includegraphics[width=8cm]{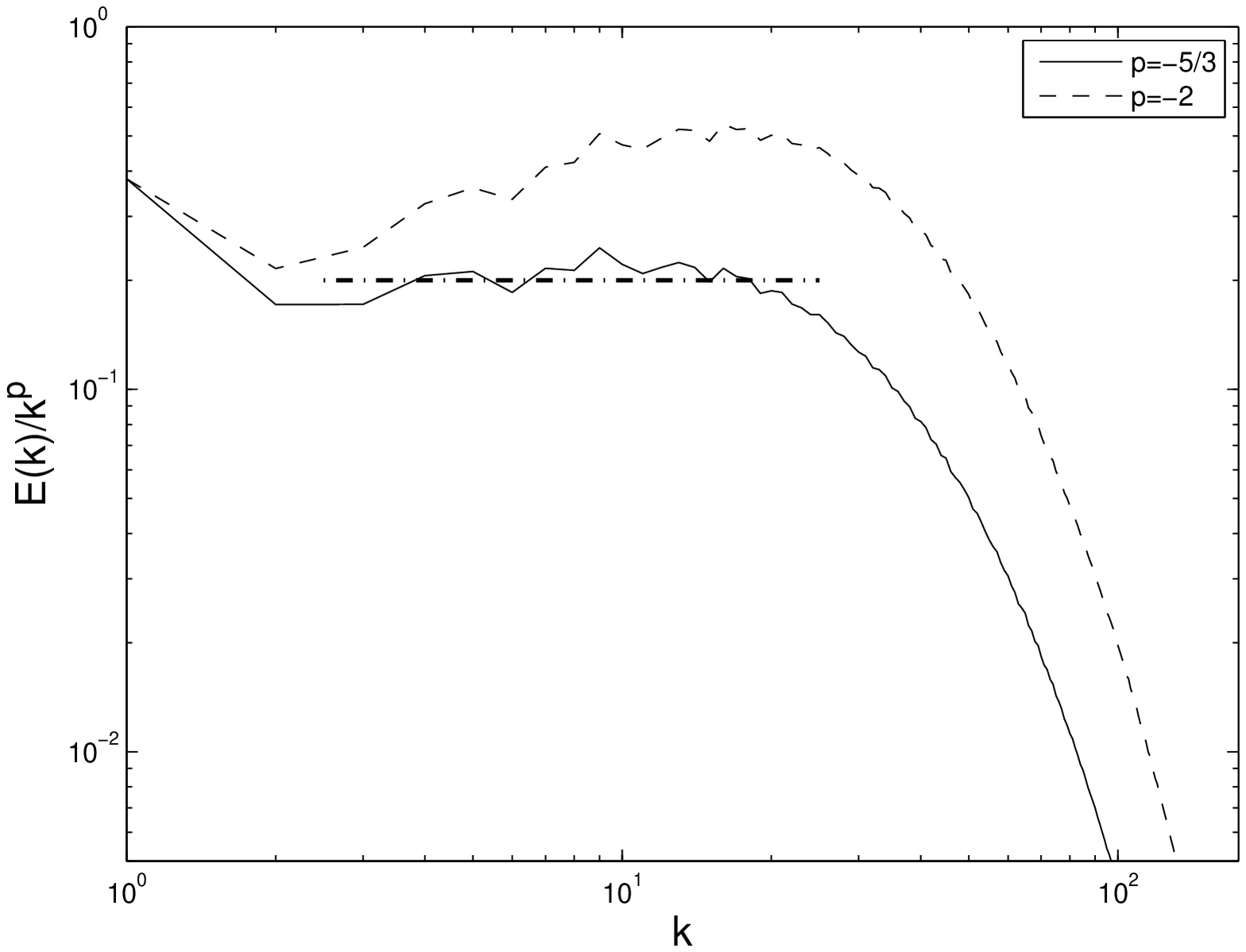} 
\caption{
$E(k)/k^p$, where $E(k)$ is the kinetic energy spectrum, averaged for one eddy turn-over time after the peak of enstrophy, and where $p$ is chosen to compensate for either a wave turbulence law ($p=-2$, dashed line) or a Kolmogorov law ($p=-5/3$, solid line); $k$ is the isotropic wavenumber. Runs were computed on grids of $512^3$ points with initial conditions at $k_0\in [1,2]$, and Reynolds numbers $Re\approx 10000$.  The dash-dotted line indicates the best fit in the inertial range. {\it Left:} $N/f=2.99, {\cal R_R}=ReRo^2\approx 3.84,  {\cal R_B}=ReFr^2 \approx 0.43$, with $Fr\approx 0.0063$ and $Ro\approx 0.019$. {\it Right:}  $N/f= 4.0, ReRo^2\approx 1749, ReFr^2 \approx 109$, with $Fr\approx 0.1$ and $Ro\approx 0.4$. Note the steeper spectrum for moderate $N/f$ and low ${\cal R_{B,R}}$, and a scaling close to a Kolmogorov law for larger $N/f$ and substantially larger ${\cal R_{B,R}}$.}
\label{spect} \end{figure*}

We integrate the incompressible Boussinesq equations in the rotating frame, with constant (solid body) rotation $\Omega$ and gravity $g$, anti-aligned in the vertical ($z$) direction, with ${ \theta}$ the buoyancy (in units of velocity), $w$ the vertical velocity, $P$ the pressure, $\nu$ the viscosity, and $\kappa$ the diffusivity:
\begin{eqnarray} 
\partial _t {\mathbf u} +{\mathbf u} \cdot \nabla {\mathbf u}  - \nu \Delta {\mathbf u} &=&  \hskip-0.07truein -\nabla P - N \theta {\bm e_z} - 2 \Omega {\bm e_z} \times {\mathbf u}  ,  \\
 \label{eq:mom} 
\partial _t \theta +{\mathbf u} \cdot \nabla \theta -  \kappa \Delta \theta &=&N w  , \\
 \nabla \cdot {\bf u}&=&0 \ .
\label{eq:temp} \end{eqnarray}
\noindent We write ${\bf u}=(u,v,w)$ and we take a unit Prandtl number, $\nu=\kappa$. The Brunt-V\"ais\"al\"a frequency is $N=[-g \partial_z \bar \theta/\theta]^{1/2}$ where {${\bar \theta}$} is the background imposed stratification. In the general case, one has inertia-gravity waves of frequency 
$$\omega_{IG}= k^{-1}\sqrt{N^2k_{\perp}^2+f^2k_z^2}$$ 
with $f=2\Omega$ (see, e.g., \cite{bartello1, nappo}).

\begin{figure}
\includegraphics[width=8.5cm]{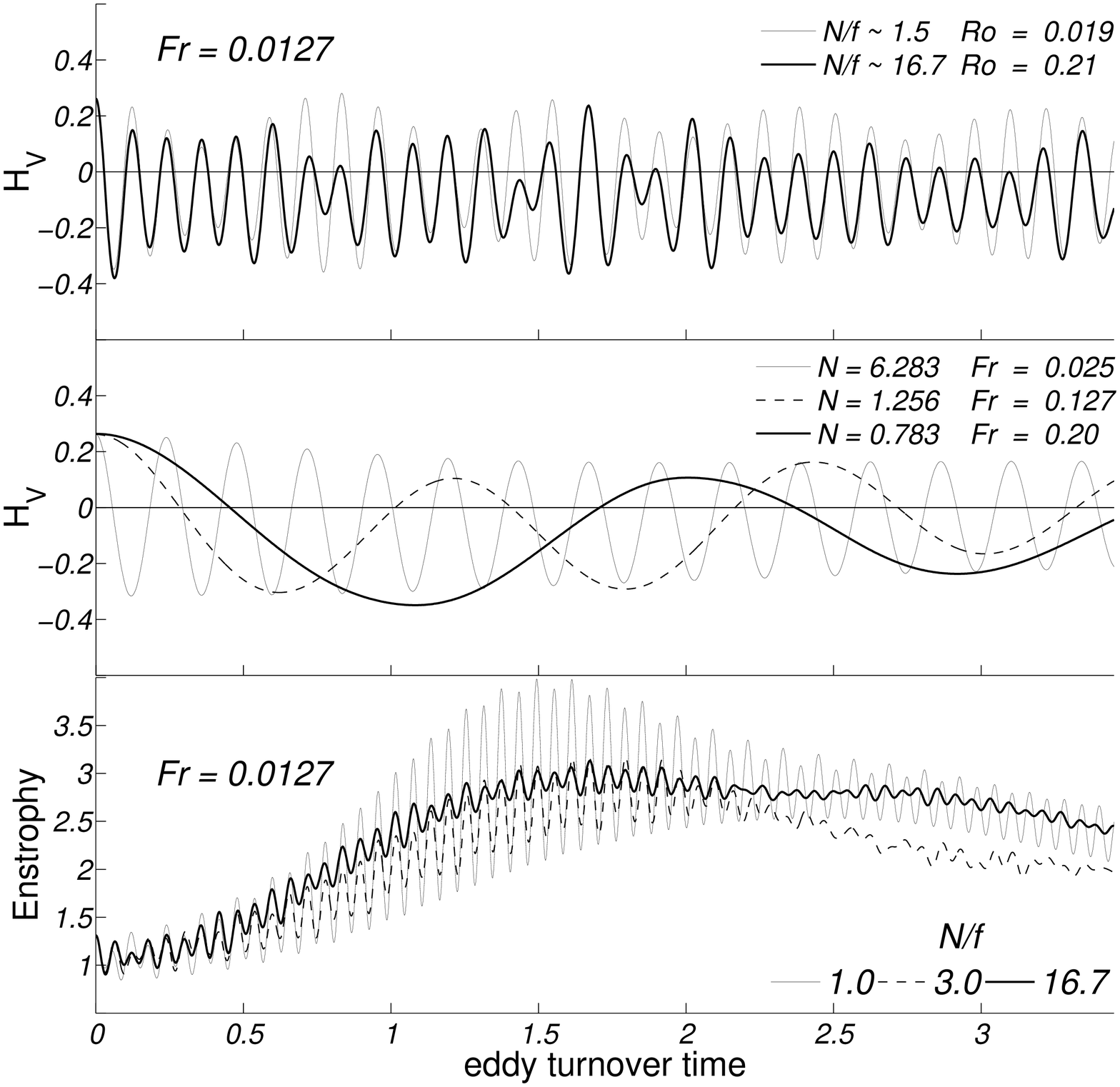}
\caption{
{\it Top and middle:} Temporal evolution of the total helicity $H_V$ in several runs with different values of $Fr$, $Ro$, and $N$, as given by the labels. Note that the time averaged value of $H_V$ is negative, indicating negative helicity prevails in these runs even when the initial value of the helicity is positive. On top are runs with the same $Fr$ whereas in the middle, runs with $N/f=1$ but with different $Fr$ are shown.
{\it Bottom:} Time evolution of the kinetic enstrophy $Z_V$ in runs with $Fr\approx 0.01$ and $N=12.56$, and with different values of $Ro$. In all panels, oscillations are due to gravity waves, with their period proportional to $N$.}
\label{temp} \end{figure}

\begin{figure*}
\includegraphics[width=8.5cm]{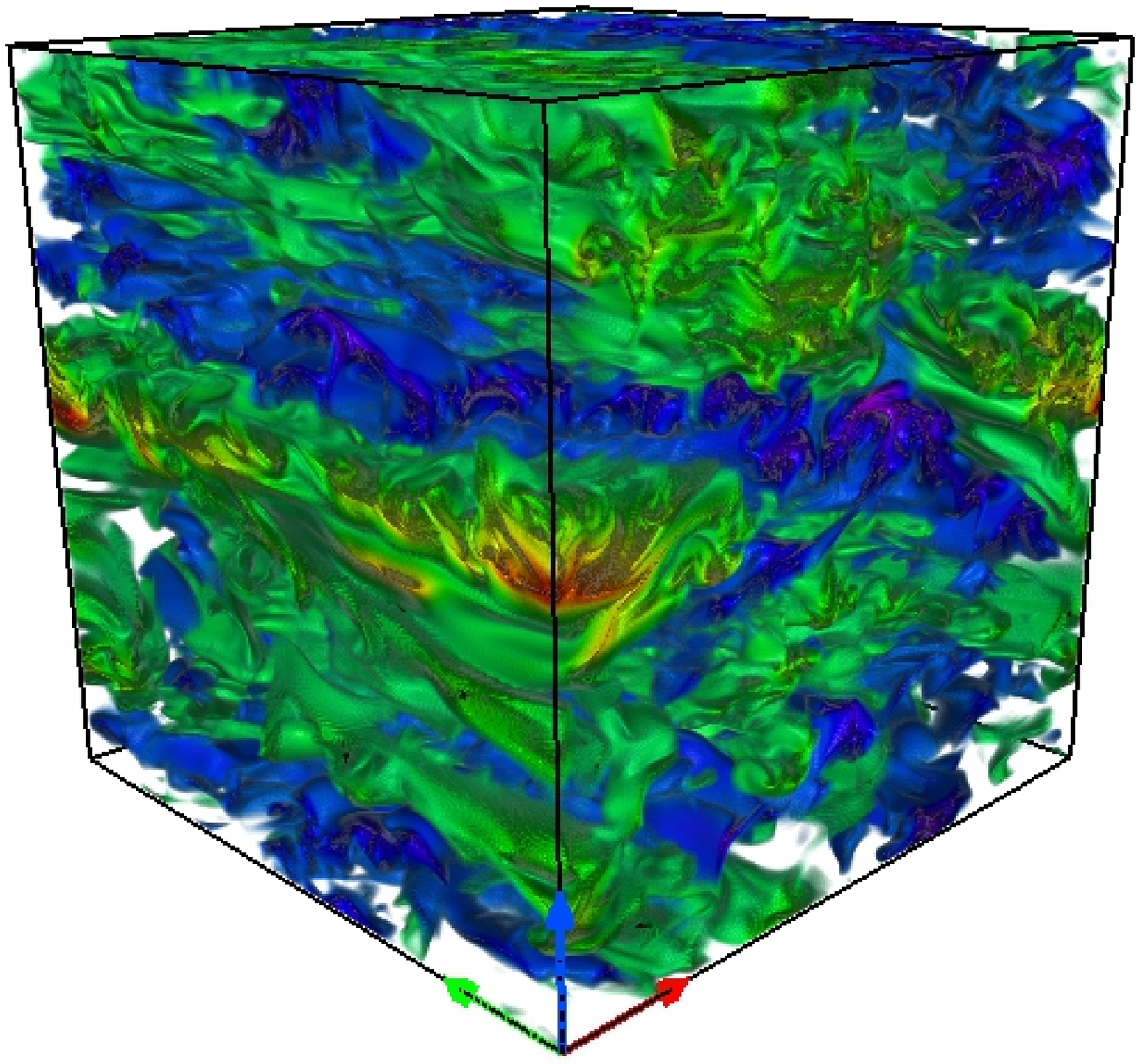}    
\includegraphics[width=9.3cm]{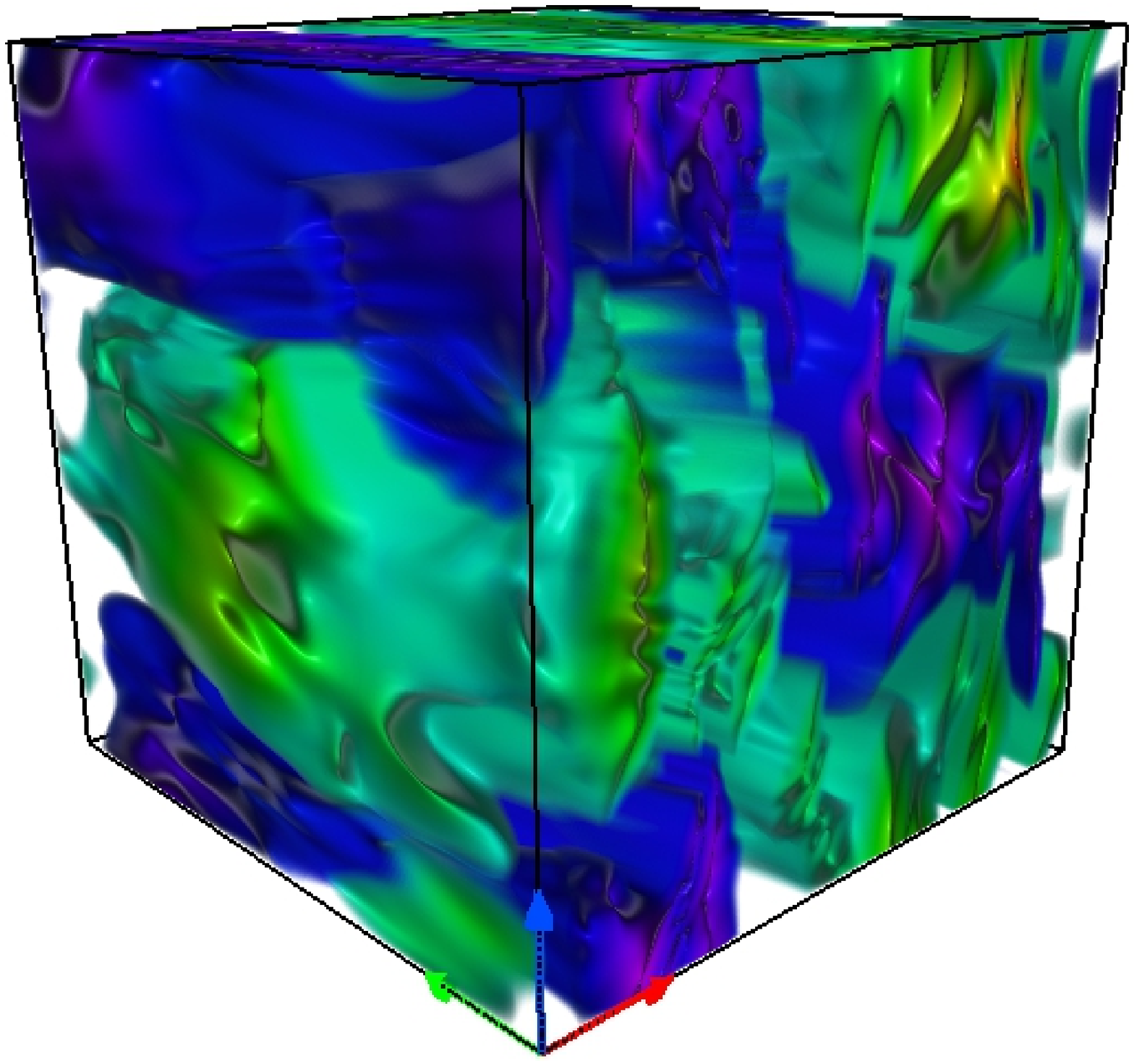}
\caption{(Color online) Visualization of the buoyancy $\theta$ in runs with $512^3$ grids, for $Re \approx10000$, $Fr = 0.1$, and $Ro$ = 0.4 (left) and for the same $Re$, $Fr = 0.025$, and $Ro = 0.05$ (right). The vertical direction is indicated by the blue arrow; dark (blue) and light (green) strata represent respectively positive and negative variations in $\theta$ around its mean, with sizable fluctuations and structuring, and with more turbulent eddies at higher Froude number.} 
\label{render} \end{figure*}

The Froude, Rossby and Reynolds numbers are defined, respectively, as 
$$Fr=\frac{u_{0}}{N L_{0}}\ , \ Ro= \frac{u_{0}}{f L_{0}} \ , \  Re=\frac{u_{0}L_{0}}{\nu} \ , $$
with $u_{0}=1$ and $L_{0}=2\pi/k_0$ respectively the r.m.s.~velocity and the scale of the initial conditions. These parameters vary in the range $0.0063 \le Fr \le 0.2$, $0.0063 \le Ro \le 3.24$, and $Re\approx 4000$ for grids of $256^3$ points, while $Re \approx10000$ using $512^3$ points. Decay is left to occur for {3.6 to 7.2} turn-over times, $\tau_{NL}=L_{0}/u_{0}$. The initial velocity field is random, with all three components non-zero, and it is centered around  wavenumbers $k_0=[1,2]$. At $t=0$, $\theta=0$, and $H_V \approx +0.2$. Other initial values have been used as well to ascertain that the results are insensitive to them. Note that we do not attempt to take initially a balanced flow; for the time-stepping point of view, there is no need to do so, since the resolutions we employ are high enough that the relatively small Froude and Rossby numbers we simulate can be handled with an explicit time stepping resolving the smallest eddy-turnover time and the smallest Brunt-V\"ais\"al\"a, inertial, and inertia-gravity frequencies. Furthermore, the generation of gravity waves that compete with turbulent eddies is part of the overall dynamics of such flows as the Reynolds number 
increases.

In the ideal ($\nu=0$) case, potential vorticity
$$PV=-fN+ f\partial_z \theta-N\omega_z+ {\bm \omega} \cdot \nabla \theta $$
is a point-wise invariant, and the total (kinetic plus potential) energy $E_T=E_V+E_P$ is conserved as well {(with $E_P= \frac{1}{2}\left<\theta^2\right>$)}, with respective enstrophies (proportional to dissipation when $\nu\not=0$ and $\kappa\not=0$), 
$$Z_V=\left<\omega^2 \right> \ , \ Z_{P}=\left< |\nabla\theta |^2 \right> .$$
Note that PV is quadratic and thus its $L_2$ norm is not conserved in general by the truncation of Fourier space used in any spectral method; however, the nonlinear term ${\bm \omega} \cdot \nabla \theta$ can be neglected in the presence of strong rotation and stratification \cite{aluie}, resulting in a quantity whose $L_2$ norm is conserved after truncation of Fourier space.

\subsection{The GHOST code and the runs}

The numerical simulations have been carried out using the Geophysical High-Order Suite for Turbulence (GHOST) code. GHOST is a pseudo-spectral framework that hosts a variety of partial differential equation (PDE) solvers optimized for studying turbulence in a $[0,2\pi]^3$ tri-periodic box, and with $2^{nd}$ or $4^{th}$-order explicit Runge-Kutta time stepping schemes. Using a cubic box and an explicit time stepping method allows in principle, given the parameters are right, for resolving all scales including the Ozmidov scale and beyond, when isotropy recovers (see, e.g., \cite{3072} for the purely rotating case). A classical 2/3 de-aliasing rule is used, meaning that for a given resolution of $n_p$ points per dimension, the maximum available wavenumber is $k_{max}=n_p/3$. The code uses a hybrid MPI/OpenMP parallelization scheme \cite{hybrid} (MPI is the Message Passing Interface library, and OpenMP stands for Open Multiprocessing, an interface to program shared memory environments). The code also has a third level of parallelization with the recent addition of support for Graphic Processing Units (GPUs) and accelerators for the Fast Fourier Transforms (FFTs). Note that the MPI communication required to complete the multidimensional Fourier transforms is all-to-all. The code uses a ``slab'' (1D) domain decomposition among MPI tasks, and OpenMP provides a second level of parallelization within each slab or MPI task. The code can compute in double or single precision based on resolution. GHOST performance has been tested on a variety of platforms, and has been shown to scale linearly up to 98304 processors, with grids up to 6144$^3$ points. Data is stored at regular intervals and post-processed, both for quantitative analysis and visualization, the latter being performed using the VAPOR visualization software \cite{clyne}.

In Table \ref{tab1} we give the major parameters of the simulations used in this paper. Note that we have restricted our analysis to moderate values of $N/f$, in particular we have for all cases $N/f \ge 1/2$. This is because, in the purely rotating case ($N\rightarrow 0$), helicity is exactly conserved and thus as one goes into that parameter regime, the creation of helicity has to become negligible with decreasing $N$ at fixed $f$; furthermore, many geophysical flows are dominated by gravity waves except at the largest scales. There are studies that show for example that, for purely rotating flows, a turbulence regime affected by waves develops for $Ro<0.2$, whereas at intermediate Rossby numbers nonlinear transfer is reduced but the inverse cascade characteristic of the bi-dimensionalization of the flow does not take place \cite{bourouiba}. Also, for  strong waves (strong rotation or stratification), turbulence barely develops resulting in steep spectra; this is related to the value achieved by the so-called buoyancy Reynolds number ${\cal R_B}$ defined below, and the equivalent concept for rotating flows, ${\cal R_R}$. Considering this region of interest in parameter space, and given the constraints of computing in three dimensions without resorting to modeling of the small-scales, only a limited exploration of the parameters is performed.

\begin{table} 
\begin{ruledtabular} \begin{tabular}{ccccccccccccc}
 $n_R$ & $Fr$& $Ro$ & in  \\
\hline 
            &                                      \\
            &     $n_p  = 256^3$ & ; &$ Re = 4189$  \\ 
            &                                      \\ 
\hline       
            1   & $0.0063$ & $0.0063$  & * , ** \\  
            2   & $0.0063$ & $0.0127$  & * , ** \\
            3   & $0.0063$ & $0.0190$  & --     \\  
            4   & $0.0063$ & $0.0507$  & --     \\   
            5   & $0.0063$ & $0.0728$  & **     \\      
            6   & $0.0063$ & $0.1013$  & * , ** \\ 
\hline
\hline 
            7   & $0.0084$ & $0.0084$  & * , ** \\   
\hline
\hline    
            8   & $0.0127$ & $0.0063$  & * , ** \\      
            9   & $0.0127$ & $0.0127$  & * , ** \\
           10   & $0.0127$ & $0.0190$  & * , ** \\       
           11   & $0.0127$ & $0.0253$  & * , ** \\  
           12   & $0.0127$ & $0.0317$  & **     \\       
           13   & $0.0127$ & $0.0384$  & **     \\       
           14   & $0.0127$ & $0.0507$  & --     \\  
           15   & $0.0127$ & $0.1013$  & --     \\      
           16   & $0.0127$ & $0.1458$  & --     \\       
           17   & $0.0127$ & $0.2111$  & * , ** \\  
\hline
\hline     
           18   & $0.0253$ & $0.0253$  & --     \\
           19   & $0.0253$ & $0.0507$  & --     \\ 
           20   & $0.0253$ & $0.1013$  & --     \\       
           21   & $0.0253$ & $0.2026$  & --     \\  
           22   & $0.0253$ & $0.2913$  & * , ** \\       
           23   & $0.0253$ & $0.4054$  & --     \\      
\hline
\hline    
           24   & $0.0507$ & $0.0507$  & --     \\     
           25   & $0.0507$ & $0.1267$  & --     \\
\hline
\hline 
           26   & $0.1013$ & $0.4224$  & --     \\      
           27   & $0.1013$ & $0.8444$  & * , ** \\      
           28   & $0.1013$ & $1.1515$  & *      \\     
           29   & $0.1013$ & $1.6888$  & *      \\
\hline
\hline  
           30   & $0.1266$ & $0.1266$  & --     \\   
\hline
\hline      
           31   & $0.2026$ & $0.2026$  & --     \\   
           32   & $0.2026$ & $0.6079$  & --     \\   
           33   & $0.2026$ & $0.8106$  & --     \\      
           34   & $0.2026$ & $1.6888$  & --     \\       
           35   & $0.2026$ & $2.3268$  & * , ** \\     
           36   & $0.2026$ & $3.2428$  & --     \\
\hline
\hline 
            &                                      \\
            &     $n_p  = 512^3$ & ;&$ Re = 10649$  \\ 
            &                                      \\
\hline       
           37   & $0.0063$ & $0.0127$  & --     \\   
           38   & $0.0063$ & $0.0190$  & *      \\  
\hline
\hline    
           39   & $0.0127$ & $0.0190$  & * , ** \\ 
           40   & $0.0127$ & $0.0317$  & * , ** \\ 
           41   & $0.0127$ & $0.0443$  &     ** \\ 
           42   & $0.0127$ & $0.0633$  & * , ** \\   
           43   & $0.0127$ & $0.1013$  & --     \\  
\hline
\hline     
           44   & $0.0253$ & $0.0507$  & * , ** \\  
\hline
\hline
           45   & $0.1013$ & $0.4053$  & *      \\       
\end{tabular} \end{ruledtabular} 
\caption{
List of runs analyzed in this paper with some characteristic parameters: run number $n_R$, linear resolution $n_p$, Reynolds $Re$, Froude $Fr$ and Rossby $Ro$ numbers. A star in the ``in'' column indicates points that are in the scatter plot with $N/f<3$, and two stars indicate those in the plot with ${\cal R_B}<20$ or ${\cal R_R}<20$ (see Figs.~\ref{merged} and \ref{scatter_inset}).}
\label{tab1}
\end{table}

\section{Results}
\subsection{\ADD{Generation of helicity for small nonlinearity}}

As mentioned in the introduction, helicity is not conserved in a rotating and stratified flow, and thus helicity can in principle be created by the flow evolution. In this section we briefly show how a balance of the forces at large scales can result in net helicity of a preferred sign in the flow. We start from the primitive Boussinesq equations given above and simplify them using several hypotheses. Assuming stationarity, weak nonlinearities and small dissipation at large scales, it results that the equilibrium level of helicity in rotating stratified turbulence is proportional to $N/f$ and to the correlation between buoyancy and vertical shear. A result consistent with this behavior was originally obtained by Hide \cite{moffatt_bk}.

\ADD{We start with the momentum equation, Eq.~(\ref{eq:mom}). As later we will compute time averages, we will assume the system is in a steady state and neglect the time derivative. We will also consider viscous effects are small, and neglect the dissipative term. Computing the vertical derivative of the remaining terms and taking the dot product of the result with ${\mathbf u}$, we get
\begin{equation}
{\mathbf u} \cdot \partial_z \left( {\mathbf u} \cdot \nabla {\mathbf u} 
  \right) = - {\mathbf u} \cdot \nabla \partial_z P - N w \theta - 2 
  {\mathbf u} \cdot \partial_z \left( \Omega {\bm e_z} \times {\mathbf u} 
  \right) ,  \\
\label{dzmom}
\end{equation}
where the velocity field ${\mathbf u}$ was written with Cartesian components $(u,v,w)$.}

\ADD{The last term in this equation is
\begin{equation}
 2 {\mathbf u} \cdot \partial_z \left( \Omega {\bm e_z} \times 
    {\mathbf u} \right) = -f (u \partial_z v - v \partial_z u) = 
    f H_\perp,
\label{coriolis}
\end{equation}
where $H_\perp$ is part of the total helicity density. Indeed, we decompose the helicity as $H_V \equiv \left<H_{\perp}\right>+\left<H_+\right>$, where the brackets denote an average, and where $H_\perp$ is the helicity density associated with ${\bf u}_{\perp}$,
\begin{equation}
H_{\perp} \equiv {\bm u_{\perp}} \cdot (\nabla \times {\bm u}_{\perp}) ,
\end{equation}
and $H_+$ is the remainder, $H_+= u\partial_yw-v\partial_xw+w\omega_z$.} With strong rotation and stratification, $ H_{\perp} \gg H_+ $, and $H_{\perp}$ alone essentially determines the total helicity. For example, measurements of $\left<H_{\perp}\right>_{\perp,z}$ (where the subindices $\perp$ and $z$ indicates the averages are volume averages performed in the horizontal and vertical directions) found in modeling simulations of hurricanes are seen to be two orders of magnitude larger than $\left<H_+\right>$ \cite{tan}. Note also that the $H_{\perp}$ density is proportional to the so-called (cell-relative) environmental helicity, when integrated over the vertical (see, e.g., \cite{marko}).

\ADD{From Eqs.~(\ref{dzmom}) and (\ref{coriolis}), it follows that:
\begin{equation}
H_\perp = -\frac{1}{f} \left[ N w \partial_z \theta + 
    {\bf u} \cdot \nabla \partial_z P + 
    {\mathbf u} \cdot \partial_z \left({\mathbf u} \cdot \nabla {\mathbf u} 
    \right) \right] .
\end{equation}
When integrated over volume, the second term vanishes for an incompressible flow. The third term is cubic in the velocity and in a turbulent flow proportional to $\epsilon (f L_z)^{-1}$, where $\epsilon$ is the energy flux, and $L_z$ a characteristic vertical scale. For flows with strong rotation and stratification, this quantity is expected to be small (see below). As a result, after integration and neglecting the third term, we obtain
\begin{equation}
\left<H_{\perp} \right>_{\perp,z} = - \frac{N}{f} \left< w 
    \frac{\partial \theta}{\partial z} \right>_{\perp,z} \ . 
\label{moffatt2}
\end{equation}
}

\ADD{This expression was obtained before by Hide \cite{moffatt_bk}, in a slightly different form after integrating by parts and assuming periodic boundary conditions
\begin{equation}
\left<H_{\perp} \right>_{\perp,z} = \frac{N}{f} \left< \theta
    \frac{\partial w}{\partial z} \right>_{\perp,z} \ . 
\label{moffatt2bis}
\end{equation}
It should be noted that the original derivation in \cite{moffatt_bk} assumes the nonlinear term is zero and that the flow is in  geostrophic balance. In that case, from hydrostatic balance $\partial_z w=0$ and helicity in the flow vanishes. Small nonlinearities are crucial to ensure that the second-order correlator in Eqs.~(\ref{moffatt2}) or (\ref{moffatt2bis}) is non-zero.}

We thus conclude that, \ADD{if nonlinearities are small}, the production of helicity in strongly rotating stratified turbulence is such that its equilibrium level is directly proportional to $N/f$, and results from \ADD{an interplay} between rotation and stratification. In the limit of $f\rightarrow \infty$ (no stratification), helicity is exactly conserved. In the limit of $N \rightarrow \infty$, stratification dominates and the evolution of helicity can only be governed by the nonlinear terms, the buoyancy and the dissipation \cite{cecilia}. Indeed, in that case dissipation is known to play a role in the overall dynamics, e.g., in the changes of potential vorticity once gravity waves start to break \cite{MCI_90}. Finally, it is interesting that $N/f$ scaling has also been advocated, for example, in the context of statistical mechanics of non-dissipative geophysical flows \cite{note2}. 

\subsection{Nonlinear effects}

\ADD{Small nonlinearities and negligible dissipation are just the beginning of the story,} with these assumptions broken when overturning takes place. For example, it is known that in three-dimensional turbulence without waves, the rate of energy dissipation can be evaluated phenomenologically as $\epsilon \sim U_0^3/L_0$, no matter how high the Reynolds number; this has been demonstrated using highly-resolved direct numerical simulations \cite{kaneda} up to grids of $4096^3$ points (for the case of a coupling to a magnetic field, in which case Alfv\'en waves are present and interact with the flow, see \cite{politano} in two dimensions (2D), and \cite{mininni_MHD} in 3D).

Similarly, there is a vast literature which concerns itself with the weak non-linear coupling of waves, e.g., through resonant interactions \cite{hassel}, through weak turbulence theory (see \cite{galtier_rot} for the rotating case, and \cite{caillol} for the stratified case), through turbulence closures \cite{jacquin_89, bellet, bourouiba}, and more recently through asymptotic approaches \cite{embid_98, julien, klein_10}. In all cases, when the rotation or stratification is not strong enough, and/or when the Reynolds number is high enough, a situation described by both the buoyancy and what can be called the inertial Reynolds number
\begin{equation}
{\cal R_B}=ReFr^2, \ \ {\cal R_R}=ReRo^2 ,
\label{RB} \end{equation}
being large enough, nonlinear couplings will take place between eddies and waves, sufficiently so that the energy will be transferred to small scales in a self-similar manner. Let us note here that in the following (see also Table \ref{tab1}), ${\cal R_{B,R}}$ are evaluated at the peak of enstrophy, using dynamical variables, i.e., based on the so-called integral scale $L_{int}=\int [E_V(k)/k]/E_V$ .

In these cases, that we will broadly call wave turbulence, the scaling laws for either rotating or  stratified flows have been deduced both phenomenologically and analytically in the framework of the aforementioned closures and theories, and a continuous power-law spectrum is expected, steeper than a Kolmogorov spectrum, because of the weakening of interactions in the presence of waves. \ADD{The energy flux $\epsilon$ is also typically reduced, in general by a factor that is proportional to the ratio of the eddy turnover time to the timescale of the waves.} Thus, we are in the presence of turbulence, but not the classical Kolmogorov turbulence, rather a wave turbulence regime that breaks down at small scales (beyond the Ozmidov scale). Two typical cases of the isotropic energy spectra are shown in Fig. \ref{spect}. We observe well developed kinetic energy spectra, rather steep for $N/f<3$ ($E(k)\sim k^{-2}$), whereas the power law is closer to a Kolmogorov law for $N/f=4$. In both cases, the dissipation scale $\eta$ (evaluated using  a Kolmogorov spectrum) is barely resolved: one finds $\eta\approx 0.041$ and $k_{\eta} \approx 154$ for the case where $N/f=2.99$, and $\eta\approx 0.044$ with $k_{\eta} \approx 144$ for the case where $N/f=4.0$. Note that in the latter case, the turbulence is stronger and the spectra are not quite as well resolved, but the point of this study is not to examine Kolmogorov turbulence; in the presence of waves, one can simulate higher Reynolds numbers at a given resolution than in the absence of waves, again the set of governing parameters being  ${\cal R_{B,R}}$ rather than $Re$.
  
Note that we do not intend to perform a detailed analysis of wave-mode and vortical-mode perpendicular and parallel energy and helicity spectra here, but rather show that turbulence and helicity develop in the flows we study. There are several examples in the literature of such studies at high resolution for the energy (i.e., $E_{\perp, \parallel}(k_{\perp}, k_{\parallel})$, see for example \cite{kimura, werne, almalkie} for the purely stratified case, and \cite{3072} for  rotating turbulence). In our case, the choice of isotropic spectra is 
sufficient to show that there is indeed for these parameter regimes power-law spectra that develop through nonlinear mode coupling. These spectra may display intermittency at small scale, a phenomenon that would require substantially higher numerical resolutions to study.

\ADD{As the Reynolds numbers are increased, the amplitude of the nonlinear term can be expected to increase, and helicity should be given by
\begin{equation}
\left<H_{\perp} \right>_{\perp,z} = - \frac{N}{f} \left< w 
    \frac{\partial \theta}{\partial z} \right>_{\perp,z} - 
    \frac{1}{f} \left< {\mathbf u} \cdot \partial_z 
    \left({\mathbf u} \cdot \nabla {\mathbf u} \right) 
    \right>_{\perp,z} .
\label{correction}
\end{equation}
In the simulations presented in the following section, we compared the ratio of the second term to the first in Eq.~(\ref{correction}). For simulations with $0.5 \le N/f \le 3$ and with $Re Ro^2<20$ and $Re Fr^2<20$, the amplitude of the second term is smaller than $10-20\%$ of the first term in most of the runs, and increases for runs with larger values of $N/f$.} 

Therefore, we can expect that for very small values of the control parameters, and for flows for which the geostrophic and hydrostatic balance holds, the latter implying $\partial_z w=0$, the helicity should remain zero. As fluctuations develop, and as weak nonlinear perturbations come into play, small departures from geostrophy will develop allowing for non-zero correlations between buoyancy and vertical velocity, as appears in Eqs.~(\ref{moffatt2}) and (\ref{moffatt2bis}) and as can also be expressed in two-scale turbulence closure formalisms. In that case, we can expect helicity to develop. Helicity should be created at large scales, where rotation and stratification dominate over the nonlinear term, and may be transferred to smaller scales. \ADD{Finally, for runs with stronger nonlinearities and large values of the control parameters, we can expect deviations from the prediction in Eqs.~(\ref{moffatt2}) and (\ref{moffatt2bis}), associated with the extra term in Eq.~(\ref{correction}). A detailed study of these deviations is left for future work.}

\section{Parametric study}

We have performed nine runs on grids of $512^3$ points, and 36 runs on $256^3$ grids, up to the peak of dissipation and beyond, with similar (but not identical) random initial conditions and $N/f \in [1/2, 16.7]$.

Figure \ref{temp} gives the temporal evolution of helicity (top {and middle) for several runs at either fixed $Fr$ or fixed $Ro$}, 
and of kinetic enstrophy $Z_V$ (bottom) for several flows at fixed $Fr$; the potential enstrophy $Z_{P}$ shows a behavior similar to $Z_V$, with slightly smaller values. Note that in all quantities the oscillations are due to gravity waves because of the fact that our initial conditions are chosen to be unbalanced, and their periods are proportional to $N$. Across all runs, the maximum of $Z_V$ varies from 30 (for weak waves) to $\approx 2.5$, corresponding to the smallest Froude number considered. The time to reach this maximum varies from 1.5 to 3.2 $\tau_{NL}$. The growth of enstrophy is typical of a turbulent flow, and is due to vortex stretching. The growth in the presence of waves is weaker, a characteristic of a weak turbulence regime.

The overall structures in this type of flows are shown in Fig.~\ref{render}, which displays volume rendering of buoyancy right after the peak of enstrophy for a run with $Fr=0.1$ and $N/f=4$ (left), and for a run with $Fr=0.025$ and $N/f=2$ (right), both performed on grids of  $512^3$ points and with identical initial Reynolds numbers. The 3D rendering puts in evidence  the stratification and the presence of large-scale layers; small-scale features with curved ribbons also occur for the run with smaller stratification. The run with $Fr=0.1$ shows strong turbulent fluctuations, whereas the run with $Fr=0.025$ is smoother,  with weaker small-scale fluctuations.

\begin{figure}
\includegraphics[width=0.48\textwidth]{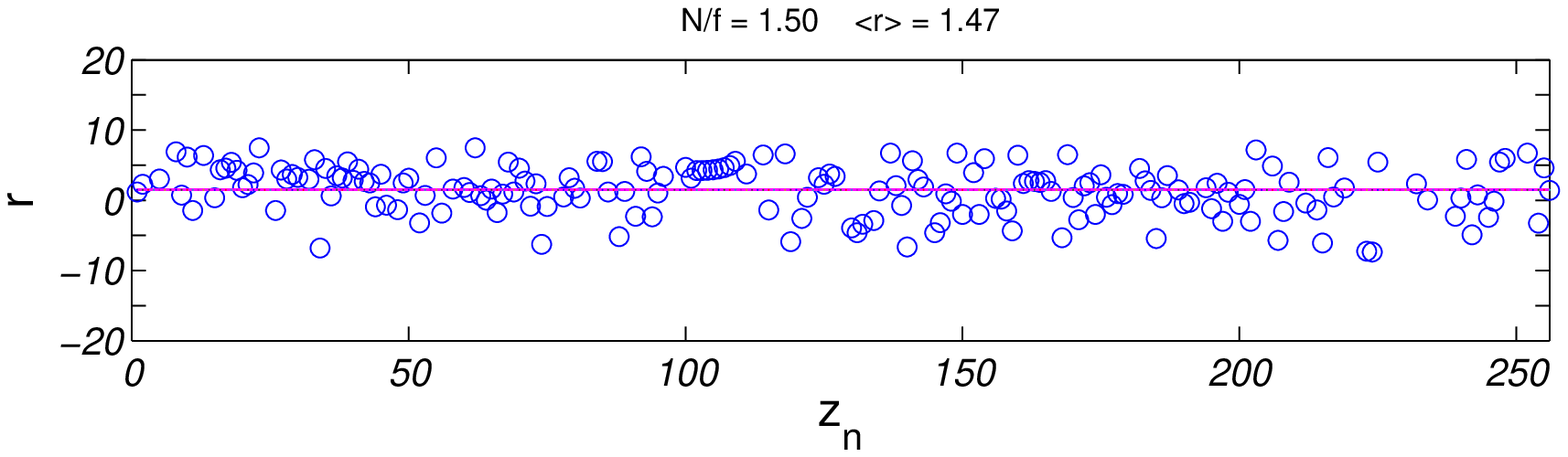} 
\includegraphics[width=0.48\textwidth]{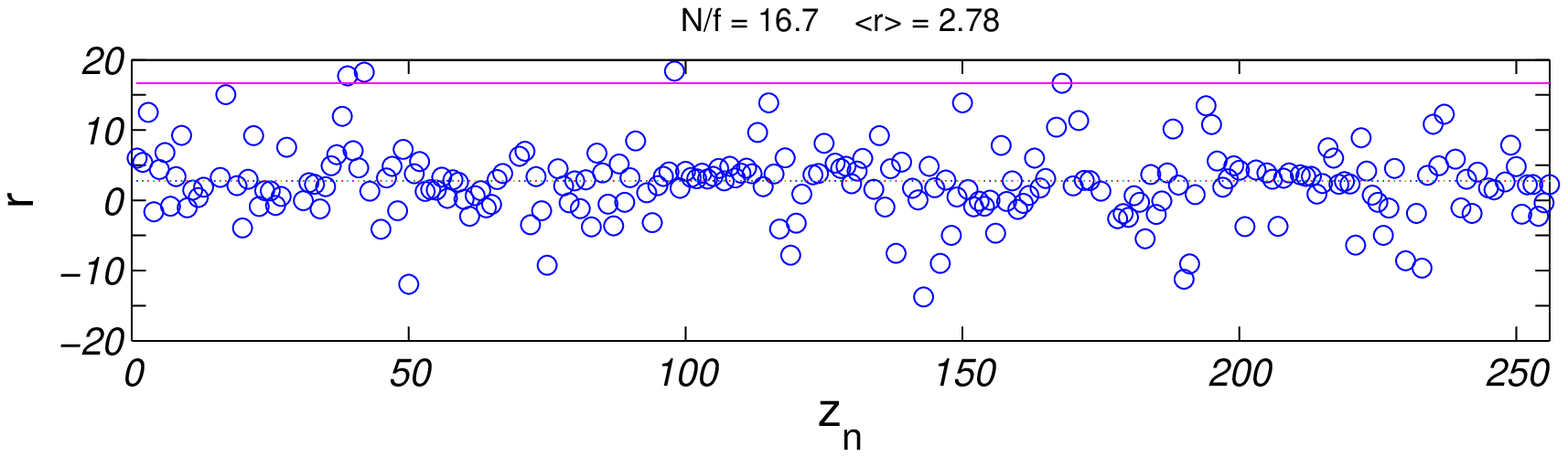} 
\caption{
Variations of $r = -\left<H_{\perp}\right>_{\perp,t}/\left<w \partial_z \theta \right>_{\perp,t}$ (see Eq.~\ref{moffatt2}) with vertical layers of index $z_n$; $n\in [1,256]$ is the index of the vertical plane, and the data is  temporally averaged around the peak of enstrophy. The horizontal lines give the geostrophic balance prediction. Both runs are performed on grids of $256^3$ points with $Re\approx 4\times 10^3, Fr=0.0127$, and $ReFr^2\approx 0.672$.
{\it Top:} $N/f=1.5, ReRo^2\approx 1.51$. 
{\it Bottom:}  $N/f= 16.7, ReRo^2\approx 186.6$. In the latter case, the prediction stemming from \ADD{assuming weak nonlinearities} no longer applies.}
\label{merged} \end{figure}

We now examine the relation given by Eq.~(\ref{moffatt2}). In Fig.~\ref{merged} is given the variation with the vertical index $z_n$ {(i.e., the vertical grid point, in runs with $n_p =256$)} of the ratio $ r = -\left< H_{\perp} \right>_{\perp, t}/\langle w \partial_z \theta \rangle_{\perp, t}$. The sub-index $t$ indicates the quantities were not only averaged in planes perpendicular to $z$, but also averaged in time over the peak of enstrophy. The ratio is shown for two runs with different values of $N/f$, namely 1.5 (top) and 16.7 (bottom); the horizontal line gives the prediction based on weak nonlinearities, i.e., $N/f$. One observes regular variations around the mean value in the vertical, so in the following we shall perform vertical averaging as well. These fluctuations are likely associated with alternating quiescent and turbulent patches where the advection term is strong. In spite of these fluctuations, the run with $N/f=1.5$ shows good agreement with the prediction based on  \ADD{weak nonlinearities}, while the run with $N/f=16.7$ does not. After performing a detailed analysis of all the runs, it is found that for all runs a good agreement with Eq. (\ref{moffatt2}) obtains for $N/f< 3$; it is also fulfilled when $ReRo^2<20$ together with $ReFr^2< 20$, 
i.e., for strong enough waves \ADD{and weak nonlinearities as explained in the previous section}. Note that the importance of the buoyancy Reynolds number has been identified previously, e.g., in the context of an emphasis on the role of anisotropy and the 
onset of Kelvin-Helmholtz instabilities due to vertical shearing \cite{riley}. 

\begin{figure}
\includegraphics[width=8.7cm]{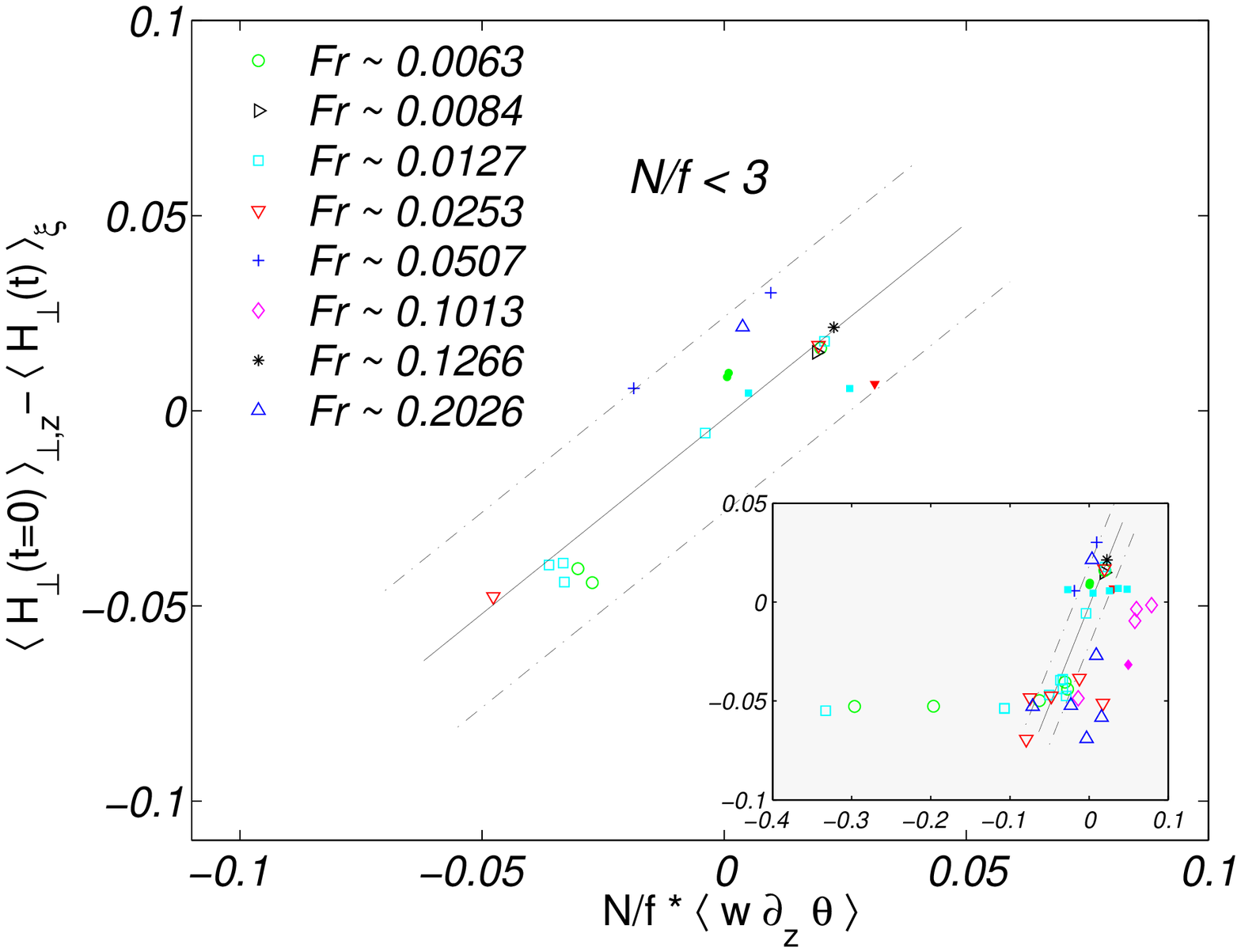} 
\includegraphics[width=8.7cm]{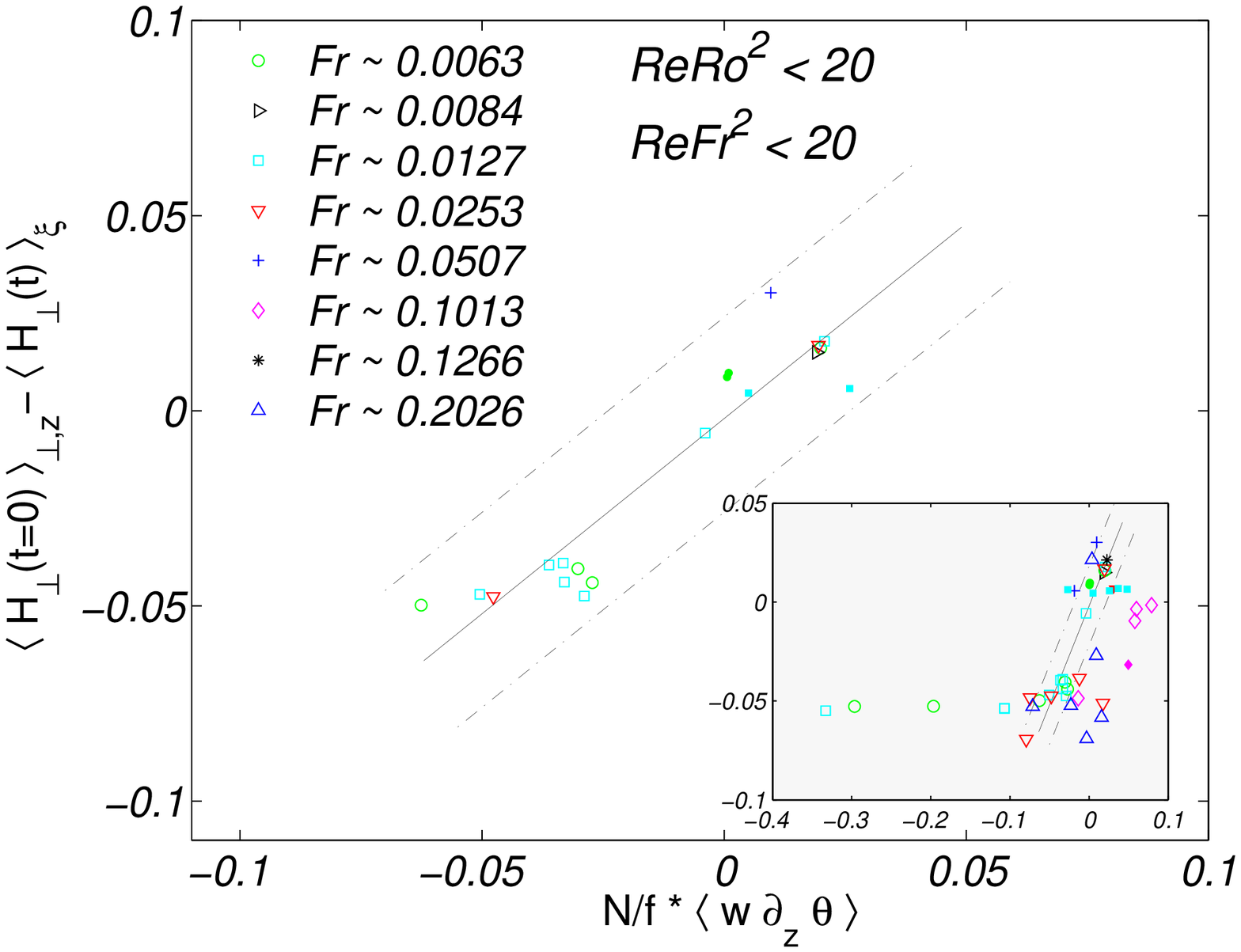} 
\includegraphics[width=8.7cm]{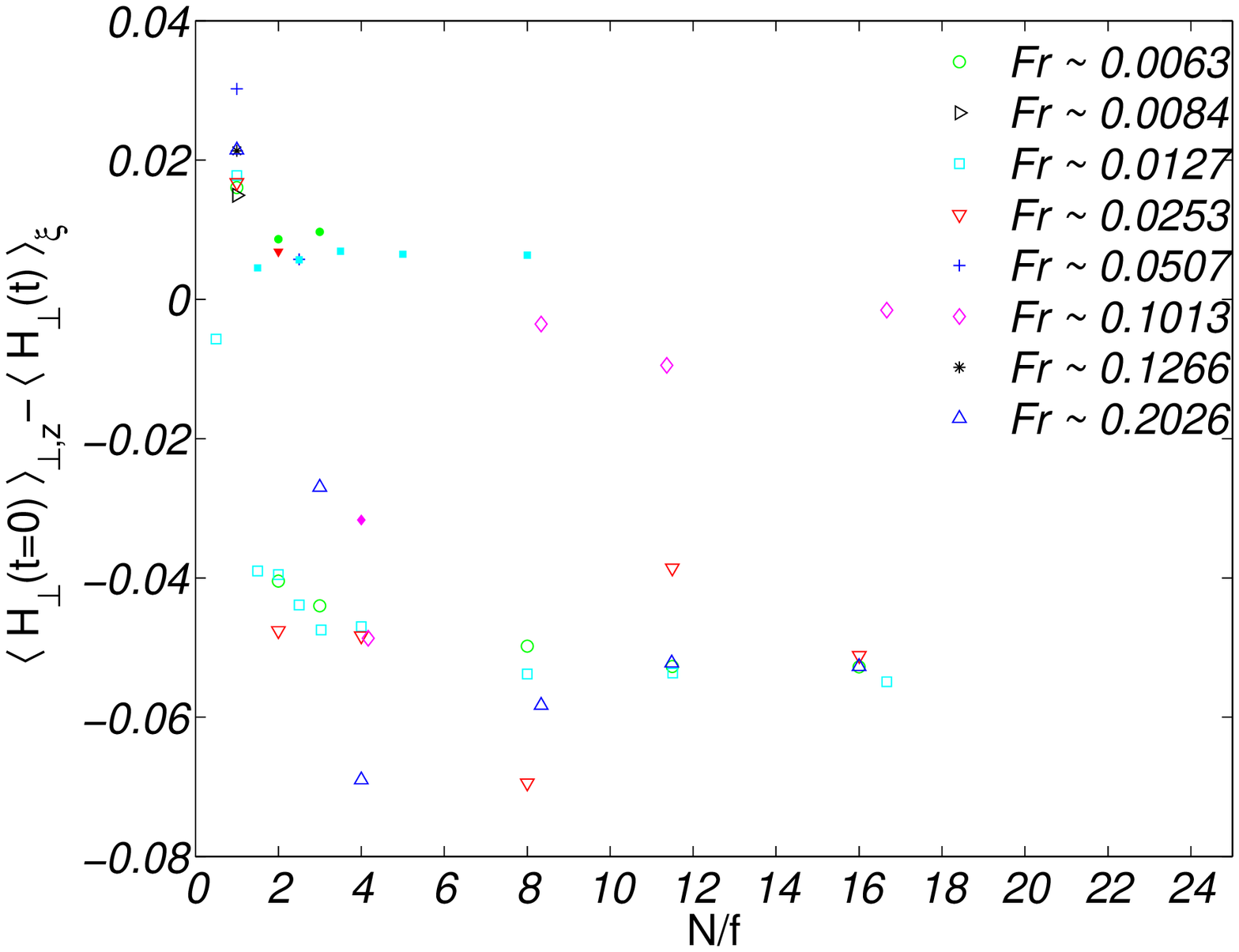} 
\caption{
Helical geostrophic balance $ \left<H_{\perp}(t=0)\right>_{\perp, z} - \left< H_{\perp}(t)\right>_{\xi}$ as predicted by Eq.~(\ref{moffatt2}), temporally and vertically averaged for runs with $N/f < 3$ (top). In the shaded insets, the same data is given for all 45 runs of this study. Each symbol corresponds to a value of $Fr$ (see labels), and the 9 solid symbols indicate runs  on grids of $512^3$ points at higher Reynolds numbers. In the middle graph, the points are selected using ${\cal R_B}<20$, ${\cal R_R}<20$ (the inset shows again all points for comparison). For completion the same scatter plot as a function of $N/f$ only, using the same  symbols, is given at the bottom.}
\label{scatter_inset} \end{figure}

Given the measurable vertical variations observed in Fig.~\ref{merged}, and following the expression in Eq.~(\ref{moffatt2}), we display  in Fig.~\ref{scatter_inset} (top) a scatter plot of $\left<H_{\perp}(t=0)\right>_{\perp,z} - \left< H_{\perp}(t)\right>_{\xi}$ as a 
function of {$N/f \left< w \partial_z \theta \right>$} for all runs with $N/f<3$; $\xi=\perp, t, z$ represents averaging on horizontal planes, for half an eddy turn-over time after the maximum of enstrophy, and over all the vertical planes as well. This allows for smoothing over temporal variations due to gravity waves, and over the vertical inhomogeneities of the flow that are inherent in strongly stratified flows as discussed before. The symbols in Fig.~\ref{scatter_inset} indicate different Froude numbers, and 
the filled symbols are used for runs on grids of $512^3$ points. For the runs with $N/f<3$, all points lie close to a straight line
with slope one, showing that the helicity created and the source of helicity according to Eq.~(\ref{moffatt2}) are linearly correlated. The result presented in the figure is robust for different choices of range over which the temporal average is performed, with windows of 0.5, 1 and 1.5 eddy turn-over times after the peak of the enstrophy. When averaging over later times, say from 3.6 to 7.2, good agreement with the prediction of helical geostrophic balance also holds, in part due to the fact that at late times, the Reynolds number has decreased and waves are now more easily predominant, with smaller Froude and Rossby numbers. It is interesting that the range of validity in $N/f$ corresponds in part to the range identified in \cite{smith_2002} on the basis of 
a lack of resonant interactions for these parameters.

We also show in the shaded insets of Fig.~\ref{scatter_inset} (top, middle) the same quantity for all the runs (i.e., all values of $N/f$, 
corresponding to all 9 runs with $512^3$ points and 36 runs with $256^3$ points). {For $N/f>3$}, creation of helicity still occurs, though not quite at the level predicted by Eq.~(\ref{moffatt2}). The middle graph in Fig. \ref{scatter_inset} gives the same scatter plot but thresholded for the buoyancy and the inertial Reynolds numbers, ${\cal R_{B,R}}<20$ (the inset shows again all points for comparison). In both cases of thresholding, about half the points are selected (roughly, 20), and the points in common between 
the top and middle scatter plot are 80\% that again (namely, 16, see Table \ref{tab1}). We can interpret this fact by saying that scales (temporal and spatial) are not sufficiently separated and it is difficult to sort out what may be the dominant effect for flows to obey the relationship of Eq.~(\ref{moffatt2}): a comparable rotation and stratification, or simply a low buoyancy Reynolds number. Finally, for completion, we give in Fig.~\ref{scatter_inset} (bottom) $\left<H_{\perp}(t=0)\right>_\perp - \left< H_{\perp}(t)\right>_{\xi}$ as a function of $N/f$ alone. Note how most of the points pile up near negative values, even for large values of $N/f$. Since a growth of net  helicity, and similarly of relative helicity,  is not observed in freely decaying 3D homogeneous turbulence, with no 
rotation and no stratification, this further confirms that the production of helicity is characteristic of the regime under study. We
finally note that it seems to be controlled more by the imposed stratification than by the Rossby number, in agreement with the 
fact that in rotating turbulence, helicity is conserved in the absence of dissipation.

\section{Concluding remarks}

A parametric study of decaying rotating stratified turbulence shows that helicity is spontaneously produced at large scales, and that for $N/f< 3$ (or, $ReFr^2 < 20$ together with $ReRo^2 < 20$), its value is associated with \ADD{correlations  between buoyancy and vertical shear}, as derived in \cite{moffatt_bk} (for non-stratified flows see \cite{klein_10}, and for the magnetic case see \cite{moffatt_08}). This creation of helicity still takes place for larger values of $N/f$, and thus confirms the possibility, for geophysical and astrophysical flows, that the combination of rotation and stratification creates helicity which in turn can be the source of large-scale magnetic fields, as observed in stars and planets. 

Helicity production in rotating stratified flows can also be related to the observation of large-scale helicity in the atmosphere of the Earth, although it is not occurring in our study through an instability involving anisotropic small-scale helicity as studied before in \cite{aka, levina}, but rather through a quasi-linearization of the large-scale dynamics. Such large-scale helical flows might be relevant to the persistence of large-scale convective storms and to the onset phase of hurricanes \cite{moli, levina}. It has also been shown that helical motions can be associated with the spiral rainbands of hurricanes when taking moisture into account in the dynamical equations \cite{tan}.

The observed saturation in the level of helicity for larger values of $N/f$ and for sufficiently strong stratification can likely be
understood in terms of the presence of vertically-sheared horizontal flows in that regime (see, e.g., \cite{embid_98, smith_2002}), a tendency that persists in the absence of rotation \cite{godeferd}. The generation of helicity requires an interplay between stratification and rotation, and when stratification dominates, vertical and horizontal motions are less correlated. \ADD{For large values of $ReFr^2$ and $ReRo^2$, the deviations from the prediction assuming weak nonlinearities can also be associated with an increase of the amplitude of the nonlinear term in the momentum equation, as verified in the simulations by direct estimation of the amplitude of the different terms in the equation for the helicity.}

The fact that rotating stratified flows can spontaneously produce large-scale helicity opens new lines of research and questions. For example, is there a detailed role to be played by potential vorticity conservation on the emergence of helicity? How would the inclusion of either shear, radiation, moisture, or some general forcing in Eq.~(\ref{eq:temp}) modify these results? And finally, how would turbulence affect significantly the creation of helicity, as the Reynolds numbers are further increased? Indeed, mixing is thought to have two transitions in terms of ${\cal R}_B$:  in the presence of an imposed shear, it was shown in \cite{shih_05} (see also \cite{lindborg}) that below 7, molecular diffusion is observed, with basically no turbulence; the intermediate regime $7< {\cal R}_B < 100$ follows a linear Osborn diffusion law \cite{osborn}, and above that value, a new regime is reached with diffusivity scaling as  ${\cal R}_B^{1/2}$. The latter regime is of course what matters for geophysical flows  with ${\cal R}_B\approx 10^8$, such as in the meridional overturning circulation, central to climate dynamics, but is it affected by the production of helicity?

We thus plan to pursue this study concerning the role of helicity in rotating stratified turbulence in order to help  decipher the different mechanisms at play, a study that will eventually lead to better sub-grid scale models of such flows that are needed to obtain a more accurate representation of enhanced diffusivities in weather and climate models.

\acknowledgments

We thank an anonymous referee for remarks that led to clarifications and improvements of the paper. This work was sponsored by an NSF/CMG grant, 1025183 and by an NSF cooperative agreement through the University Corporation for Atmospheric Research on behalf of the National Center for Atmospheric Research (NCAR). Computer time was provided by NSF under sponsorship of NCAR. For useful discussions we also acknowledge Alain Pumir, Alain Noullez and Cecilia Rorai.

\end{document}